\documentclass[conference,9pt]{IEEEtran}
\IEEEoverridecommandlockouts

\usepackage{cite}
\usepackage{amsmath,amssymb,amsfonts}
\usepackage{algorithmic}
\usepackage{graphicx}
\usepackage{booktabs}
\usepackage{textcomp}
\usepackage{xcolor}
\usepackage{xspace}
\usepackage{url}
\usepackage{multirow}
\usepackage{floatrow}
\def\BibTeX{{\rm B\kern-.05em{\sc i\kern-.025em b}\kern-.08em
    T\kern-.1667em\lower.7ex\hbox{E}\kern-.125emX}}
\begin{document}

\title{Revisiting Acoustic Features for Robust ASR}

\author{\IEEEauthorblockN{Muhammad A. Shah}
\IEEEauthorblockA{\textit{Language Technologies Institute} \\
\textit{Carnegie Mellon University}\\
Pittsburgh, PA, USA \\
mshah1@cmu.edu}
\and
\IEEEauthorblockN{Bhiksha Raj}
\IEEEauthorblockA{\textit{Language Technologies Institute} \\
\textit{Carnegie Mellon University}\\
Pittsburgh, PA, USA \\
bhiksha@cs.cmu.edu}
}

\maketitle
\newcommand{\logspec}{LogSpec\xspace}
\newcommand{\logmelspec}{LogMelSpec\xspace}
\newcommand{\mfcc}{MFCC\xspace}
\newcommand{\freqmask}{FreqMask\xspace}
\newcommand{\gammspec}{GammSpec\xspace}
\newcommand{\gammfreqmask}{GammFreqMask\xspace}
\newcommand{\pnc}{PNC\xspace}
\newcommand{\pncc}{PNCC\xspace}
\newcommand{\diffgammspec}{DoGSpec\xspace}
\newcommand{\mlses}{MLS-es\xspace}
\newcommand{\librispeech}{LibriSpeech\xspace}
\newcommand{\tedlium}{TEDLIUM\xspace}
\begin{abstract}
Automatic Speech Recognition (ASR) systems must be robust to the myriad types of noises present in real-world environments including environmental noise, room impulse response, special effects as well as attacks by malicious actors (adversarial attacks). Recent works seek to improve accuracy and robustness by developing novel Deep Neural Networks (DNNs) and curating diverse training datasets for them, while using relatively simple acoustic features. While this approach improves robustness to the types of noise present in the training data, it confers limited robustness against unseen noises and negligible robustness to adversarial attacks. In this paper, we revisit the approach of earlier works that developed acoustic features inspired by biological auditory perception that could be used to perform accurate and robust ASR. In contrast,  Specifically, we evaluate the ASR accuracy and robustness of several biologically inspired acoustic features. In addition to several features from prior works, such as gammatone filterbank features (\gammspec), we also propose two new acoustic features called frequency masked spectrogram (\freqmask) and difference of gammatones spectrogram (\diffgammspec) to simulate the neuro-psychological phenomena of frequency masking and lateral suppression. Experiments on diverse models and datasets show that (1) \diffgammspec achieves significantly better robustness than the highly popular log mel spectrogram (\logmelspec) with minimal accuracy degradation, and (2) \gammspec achieves better accuracy and robustness to non-adversarial noises from the Speech Robust Bench benchmark, but it is outperformed by \diffgammspec against adversarial attacks.
\end{abstract}
\section{Introduction}
\label{sec: intro}
Deep Neural Networks (DNNs) have advanced the state-of-the-art in automatic speech recognition (ASR) by leaps and bounds, and have achieved near or above human performance in certain cases. Despite this, their reliability in critical real-world applications is questionable due to their sensitivity to noise and distortions, perhaps the most pernicious of which are adversarial perturbations. Adversarial perturbations are slight distortions that, while imperceptible to humans, can alter the output of DNNs when applied to their inputs \cite{goodfellow13,qin2019imperceptible}. 

Unlike DNNs human audition is generally robust to such subtle distortions. This raises the question of whether there are some aspects of human speech perception that ASR DNNs do not model. While initial ASR systems were heavily inspired by the human auditory system in the acoustic features they used, modern DNN-based systems instead train powerful function approximations like transformers and RNNs on simple features. While this approach yields high accuracy on recognition tasks under clean settings, the performance under noisy settings, especially under adversarial attacks, is severely lacking. 

Recent works have sought to improve robustness by increasing the amount and diversity of training data the models are trained on \cite{likhomanenko2020rethinking,radford2023robust} but this approach has not appreciably improved adversarial robustness \cite{shah2024speech,olivier2023transferable}. Meanwhile, It has been found, albeit in relatively simple settings, that using more biologically plausible features results in more robust ASR \cite{stern2012hearing}. However, to the best of our knowledge, the robustness, especially to adversarial attacks, of such features has not been studied in conjunction with modern ASR models trained on large and diverse speech data.

To fill this gap, in this paper, we study the impact on transcription accuracy and robustness of using acoustic features that are more biologically plausible than those commonly used for ASR (i.e. Log Mel Spectrogram). In this connection, we consider some features proposed in prior works such as gammatone filterbank features (\gammspec), and Power Normalized (Cepstral) Coefficient (PNC(C)), as well as some novel features that we have developed namely Frequency Masked Spectrogram (\freqmask) and Difference of Gammatone Spectrogram (\diffgammspec). \freqmask leverages equations derived from psychoacoustic experiments to simulate simultaneous frequency masking, the phenomenon that causes a loud frequency to mask an adjacent quieter frequency \cite{lin2015principles}. These equations can be computationally expensive so we also develop a extremely lightweight feature, \diffgammspec, which simulates lateral suppression via only a novel filterbank, and thus is as efficient as \logmelspec. Lateral suppression causes the the neural responses corresponding to a frequency to be diminished if adjacent frequencies are also present in the signal \cite{stern2012hearing}. While these phenomena have been known and earlier models of audition have incorporated them, they have been ignored in recent works. We compare these features with Mel Frequency Cepstral Coefficients (\mfcc) and Log Mel filterbank features (\logmelspec) which have become the default features of choice in ASR systems (\logmelspec moreso) and are used even by SOTA models like Whisper \cite{radford2023robust}.

In our evaluation, we train and evaluate modern transformer-based ASR DNNs (namely Conformers \cite{gulati2020conformer}, and Branchformers \cite{peng2022branchformer}) with these features on three large and diverse speech datasets, including \librispeech \cite{panayotov2015librispeech}, \tedlium \cite{hernandez2018ted} and Spanish Multi-lingual \librispeech \cite{pratap2020mls}. Our proposed features, \freqmask and \diffgammspec significantly improve the adversarial robustness while achieving similar WER on clean data as \logmelspec. Also, models with \gammspec achieve significantly lower Word Error Rate (WER) than \logmelspec on \librispeech and similar WER on \tedlium, while achieving significantly better adversarial robustness. We also evaluated the models on non-adversarial noise from the Speech Robust Bench (SRB) \cite{shah2024speech} benchmark and found \gammspec and \diffgammspec to be the most robust features respectively.
Our study provides critical insights into the role of acoustic features in robust ASR. We show that simply using \logmelspec as the default feature is not necessarily optimal, especially when there are features that can provide better accuracy and robustness at no additional computational cost.

\section{Acoustic Features}
\label{sec: feat}
\begin{figure}
    \centering
    \includegraphics[width=\linewidth]{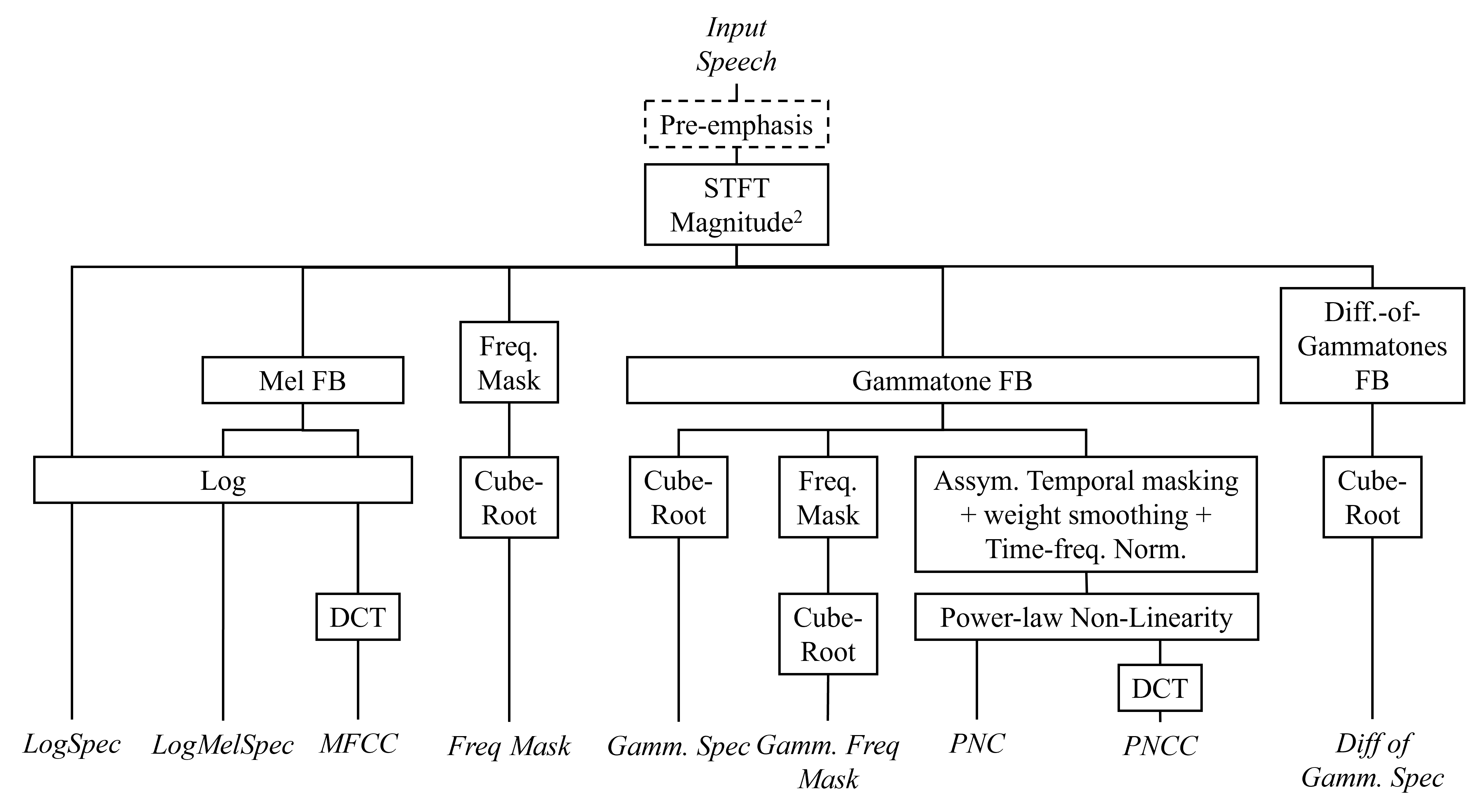}
    \caption{Overview of acoustic feature computation.}
    \label{fig:feat_tax}
\end{figure}

In this Section, we describe the acoustic features that we consider. An overview of the computations involved in these features is presented in Figure \ref{fig:feat_tax}. In this study, we only consider spectral features and thus the initial processing for all the features involves computing the audio signal's Short-Time Fourier Transform (STFT). Thereafter, the processing for each feature varies as explained below.

\textbf{Log Spectrogram (\logspec)} is obtained by applying the log nonlinearity to the STFT.

\textbf{Log Mel-Spectrogram (\logmelspec)} is obtained by applying the Mel filterbank to the STFT followed by the log nonlinearity.

\textbf{Mel Frequency Cepstral Coefficients (\mfcc)} are computed by applying the Discrete Cosine Transform (DCT) to the \logmelspec.

\textbf{Gammatone Spectrogram (\gammspec)} is obtained by applying the \textit{normalized} gammatone filterbank 
to the STFT followed by the cube root nonlinearity. The normalization divides the coefficients of each filter by their sum so that the area under each filter is 1.

\textbf{Frequency Masked Spectrogram (\freqmask)}
simulates simultaneous frequency masking, which is the phenomenon that a loud frequency (the \textit{masker}) can \textit{mask} adjacent quieter frequencies rendering them inaudible \cite{lin2015principles}. Frequency masking occurs because the presence of the masker raises the minimum level that frequencies adjacent to the masker must have to be perceptible. The minimum level behaves as a \textit{masking threshold} and frequencies with level below it are rendered inaudible.

\newcommand{\psd}{p_x(k)}
\newcommand{\npsd}[1]{\Bar{p}_x(#1)}
\newcommand{\nspsd}[1]{\Bar{p}^m_x(#1)}
\newcommand{\bi}{b(i)}
\newcommand{\bj}{b(j)}
To simulate frequency masking, we start with a STFT spectrogram, compute the masking threshold for all frequencies at each time step, and zero the energy in frequencies with level below the masking threshold. We use the method from \cite{lin2015principles,qin2019imperceptible} to estimate the masking threshold as follows. First, the log-magnitude power spectral density (PSD) is computed for each STFT frame, $x$, and frequency bin, $k$, as {$\psd=10\log_{10}\left|\frac{s_x(k)}{N}\right|^2$},
where $s_x(k)$ is the spectral magnitude in STFT bin $k$ of frame $x$, and $N$ is the window size (in samples) used to compute the STFT. 

The PSD is then normalized to have a maximum sound pressure level (SPL) of 96 dB, this is referred to as the normalized PSD, {$\npsd{k} = 96 - \max_k \{\psd\} + \psd$}.
Next, the normalized PSD is smoothed by its neighbors: {$\nspsd{k} = 10\log_10 \left[10^{\frac{\npsd{k-1}}{10}}+10^{\frac{\npsd{k}}{10}}+10^{\frac{\npsd{k+1}}{10}}\right]$}

At this stage, [CITE] performs additional processing to identify the makers. To simplify implementation we forego this step and compute the masking thresholds for all frequencies. The masking threshold induced by the masker frequency $f_i$ on frequency $f_j$ is computed as:
{$T[\bi,\bj] = \nspsd{\bi} + \Delta_m [\bi] + \text{SF}[\bi, \bj]$}, where $\bi$ is the bark scale of frequency $f_i$\footnote{frequency to bark: $b(f)=13\arctan\left(0.00076f\right)+3.5\arctan\left(f/7500\right)$. }, $\Delta_m [\bi]=-6.025 - 0.275\bi$, 
\begin{equation}
    \text{SF}[\bi, \bj] = 
    \begin{cases}
        27\Delta b_{ij},\quad \Delta b_{ij} > 0\\
        G(\bi) \cdot \Delta b_{ij}, \quad o.w
    \end{cases},\nonumber
\end{equation}
{$G(\bi) = -27 + 0.37 \max\{\nspsd{\bi}-40, 0\}$}, and {$\Delta b_{ij} = \bi - \bj$}.

Finally, for each frequency, $f_j$, we can compute the global masking threshold by combining $T[\bi,\bj]$ for all $i$ as
\begin{equation}
    \theta_x(j)=10\log_10 \left[10^{ATH(j)/10}+\sum_{i}10^{T[\bi,\bj]/10}\right],
\end{equation}
where ATH\footnote{$ATH(f)=3.64\left(10^{-3}f\right)^{-0.8} - 6.5e^{-0.6 (10^{-3}f -3.3)^2} + 10^{-15}f^4$} is the minimum PSD a frequency must have to be perceptible \textit{in quiet}.

Frequency masking is applied to the spectrogram by setting the spectral magnitude to zero at frame $x$ and frequency bin $j$ if $\nspsd{j} < \theta_x(j)$. Finally, Cube root nonlinearity is applied to the resulting spectrogram. 

\textbf{Frequency Masked Gammatone Spectrogram (\gammfreqmask)} is obtained by applying the gammatone filterbank to the STFT before applying \freqmask.

\textbf{Power Normalized Coefficients (\pnc)} \cite{kim2016power}
combine temporal masking, weight smoothing, and time-frequency normalization to achieve more robust ASR, and are computed as follows. First, the audio signal is pre-emphasized (coeff.=0.97), and the magnitude-squared spectrum is computed via STFT. Then a squared normalized gammatone filterbank, i.e. the filterbank coefficients are squared, and divided by the sum (after squaring), is applied to the spectrum to obtain the short-time spectral power, $P[m,l]$, where $m$ and $l$ represent frame and channel indices. The medium-time power is also computed as {$Q[m,l]=\frac{1}{2M+1}\sum_{i=-M}^MP[m+i,l]$}. Next, asymmetric noise suppression is applied to $Q$ to obtain $Q_{le}$. The difference between $Q$ and $Q_{le}$ is computed and half-wave linear rectified. This is followed by temporal masking, which is the phenomenon that causes an audio signal to become imperceptible if it is temporally adjacent to a louder signal. To simulate temporal masking, first the online peak power, $Q_p[m,l]$, is computed as a running maximum over $Q_0$. Then the masked signal, $R[m,l]$, is computed as 
\begin{equation}
    R[m,l]=
    \begin{cases}
        Q_0[m,l]\quad Q_0[m,l] \geq \lambda_t Q_p[m-1, l]\\
        \mu_t Q_p[m-1,l],\quad o.w
    \end{cases}\nonumber
\end{equation}
where $\lambda_t=0.85$ and $\mu_t = 2$ in \cite{kim2016power}. Next, spectral weight smoothing and mean power normalization are applied. Finally, a rate-level non-linearity (raise to power $1/15$) is applied to obtain the PNC.

\textbf{Power Normalized Cepstral Coefficients (\pnc)} \cite{kim2016power} are computed by apply the DCT to \pnc features.

\textbf{Difference of Gammatone Spectrogram (\diffgammspec)}
 is a novel feature proposed in this paper to simulate lateral suppression, the phenomenon that the response to a frequency may be suppressed if adjacent frequencies are present in the signal \cite{stern2012hearing}, even if the intensity of the latter is below the threshold of hearing. While lateral suppression has been simulated in several cochlear models in prior work \cite{lyon1984computational,slaney1988lyon}, these models are not amenable to be used in modern DNN-based systems because they tend to be computationally intensive and rather slow. As a result, lateral suppression is largely missing from modern ASR systems. To fill this gap we have developed the \diffgammspec feature for incorporating lateral suppression into DNNs. 
 
 The key component in \diffgammspec is the DoG filterbank, which is constructed as follows. First, two normalized gammatone filterbanks, $G_1$ and $G_{\alpha}$, are created such that the bandwidths of the filters in $G_1$ are scaled $\alpha$ to obtain the filters in $G_\alpha$. Next, the corresponding filters in the two filterbanks are subtracted and the resulting filters are normalized by dividing the coefficients by the sum of the positive coefficients to ensure the excitatory components sum to 1 (as in the gammatone filterbank). These operations can be formally stated as
 \begin{equation}
     G_d[i, j] = G_1[i, j] - G_\alpha[i, j]
 \end{equation}
 \begin{equation}
     \Bar{G}_d[i, j] = \frac{G_d[i, j]}{\sum_{j'=0}^{F-1} \max(G_d[i,j'],0)}
 \end{equation}
 where $i$ is the filter index, $j$ is the coefficient index, and $F$ is the number of frequency bins in the STFT. 
 The DoG filter has a suppression field over the frequencies adjacent to the filter's characteristic frequency and, thus, any non-zero energy in those frequencies will suppress the response of the filter. 
 Given an input audio signal, the \diffgammspec is computed by first applying a pre-emphasis filter, then a DoG filterbank and finally a cube-root non-linearity. 

 
\section{Experimental Setup}
\label{sec: exp}
To evaluate the robustness and accuracy of the features described in \S\ref{sec: feat}, we train various ASR models with these features on diverse datasets and then evaluate them on clean and noisy data. The details of the evaluation setup are as follows.

\subsection{Train Datasets}
We train models on three datasets, namely \librispeech \cite{panayotov2015librispeech}, \tedlium \cite{hernandez2018ted}, and the Spanish subset of Multilingual LibriSpeech (\mlses) \cite{pratap2020mls}. \librispeech and \mlses contains read speech from audio books in English and Spanish, respectively, while \tedlium contains spontaneous English speech from recorded TED talks. We use the full 960 hour and 452 hour training sets of \librispeech and \tedlium, respectively, for training. For \mlses, we use the 917 hour train set from Huggingface\footnote{\url{https://huggingface.co/datasets/facebook/multilingual_librispeech/viewer/spanish}}.

\subsection{Models}
We use recipes from SpeechBrain \cite{speechbrain} to train 13M parameter Conformer \cite{gulati2020conformer} and 104M parameter Branchformer \cite{peng2022branchformer} ASR models. The \librispeech and \mlses models use a 5k subword unigram tokenizers trained, while the \tedlium models use a 500 BPE tokenizer. The models trained on LibriSpeech also contain a transformer LM for re-scoring during decoding. During robustness evaluation, all models use beam size 10 without LM rescoring.

The original recipes use \logmelspec features so we minimally modify the recipes to incorporate the various features from \S\ref{sec: feat}. Since we set the number of filters to 80 for all filterbank-based features, we only change the \texttt{compute\_feature} field in the recipe YAML while the downstream model remains unchanged. For \logspec and \freqmask however, we do need to change the input size of the transformer because these features do not use filterbanks. 

\subsection{Evaluation Setup}
\subsubsection{Methodology and Data}
We evaluate the models' accuracy on the training dataset's official test subsets. To evaluate robustness we use Speech Robust Bench (SRB) \cite{shah2024speech}, a recently released robustness benchmark for ASR models. SRB contains multi-lingual speech with more than 100 types of noises and distortions. At a high-level, the distortions in SRB fall into 5 categories: inter-personal communication (drawn from CHiME \cite{barker2017chime} and AMI \cite{kraaij2005ami} corpora), environmental effects (environmental noise from ESC-50, MS-SNSD, MUSAN and WHAM) and room impulse responses from \cite{kinoshita2013reverb}), digital augmentations (white noise, special effects, audio processing operations like resampling, gain and filtering), speech variations (accented speech from CommonVoice \cite{ardila2019common}, and text-to-speech using YourTTS \cite{casanova2021yourtts}), and adversarial attack (untargeted PGD \cite{madry18}). While not part of SRB, we also evaluate the models on the targeted imperceptible attack \cite{qin2019imperceptible}. This attack specifically exploits frequency masking to add noise in the spectral regions that are likely to be masked by human hearing, and, thus, is a good test for features like \freqmask and \diffgammspec, which simulate frequency masking and lateral suppression. Since this attack is very computationally expensive, we evaluate only LibriSpeech models against it using 10 utterances from the LibriSpeech dataset.

\subsubsection{Metrics}
To measure accuracy we use \textit{Word Error Rate} (WER). To measure robustness we use \textit{WER Degradation} (WERD) and \textit{Normalized WERD} (NWERD) \cite{shah2024speech}. WERD is computed as the difference between the model's WER on the clean test subset of its training dataset and its WER on the noisy testing data. NWERD is computed by dividing WERD by a measure of speech quality, specifically DNSMOS \cite{reddy2020interspeech} and/or PESQ \cite{rix2001perceptual}, such that errors on less distorted utterances are penalized more than errors on more distorted utterances. This is done because a robust model should not compromise accuracy on cleaner utterances, which represent the average use case while improving accuracy on severely distorted utterances, which are rare. To evaluate robustness against the targeted imperceptible adversarial attack \cite{qin2019imperceptible}, we compute the WER between the target phrase and the predicted transcript, and the \textit{Signal-to-Noise Ratio} (SNR) of the adversarially perturbed utterance. The attack is successful if the WER is low and the SNR is high because this indicates that the attack induced the desired transcription without impacting the intelligibility of speech in the utterance.
\section{Results}
\label{sec: exp}
\subsection{Accuracy on Clean Utterances}
\begin{table}[t!]
    \centering
    \small
    \begin{tabular}{p{1.4cm}p{1.7cm}p{1.3cm}p{1.2cm}p{1.2cm}}
        Dataset/ Model & Feature & Test-clean (noLM, bs=10) & Test-clean (bs=66) & Test-other (bs=66) \\
        \midrule
         \multirow{8}{*}{\shortstack[l]{LibriSpeech/\\Conformer}} & \logmelspec & 5.25 & 2.53 & 6.04\\
        & \diffgammspec & 5.27 & 2.50 & 6.17\\
        & \freqmask & 5.71 & 2.72 & 7.85\\
        & \gammfreqmask & 5.02 & 2.46 & 6.32\\
        & \gammspec & \textbf{4.65} & \textbf{2.29} & \textbf{5.62}\\
        & \logspec & 4.66 & 2.30 & 5.82 \\
        & \mfcc & 8.31 & 3.23 & 9.44\\
        & \pncc & 8.35 & 3.42 & 9.44\\   
        & \pnc & 5.69 & 2.50 & 6.70\\
        \hline
        \multirow{3}{*}{\shortstack[l]{LibriSpeech/\\Branchformer}} & \logmelspec & 3.66 & \textbf{2.01} & \textbf{4.78}\\
        & \diffgammspec & 3.12 & 2.11 & 5.21 \\
        & \gammspec & \textbf{3.11} & 2.07 & 5.25 \\
        \midrule
        & & bs=10 & bs=66 & \\
        \midrule
        \multirow{2}{*}
        {\shortstack[l]{\tedlium/\\ Branchformer}}
         & \logmelspec & 15.07 & 7.52 & \\
        & \diffgammspec & 16.70 & 8.21 & \\
        \hline
       \multirow{2}{*}
       {\shortstack[l]{\mlses/\\
       Branchformer}} & \logmelspec & 6.20 &  6.11 & \\
        & \diffgammspec & 6.43 & 6.19 & 
    \end{tabular}
    \caption{The WER of ASR models with different features on the original test sets of \librispeech, \tedlium and \mlses with different beam sizes (bs).}
    \label{tab:clean-wer}
\end{table}
Table \ref{tab:clean-wer} shows the WER of the various models and features on the unmodified test subsets of \librispeech, \tedlium and \mlses. We observe that despite being default features of choice in modern ASR systems including models like Whisper \cite{radford2023robust} and Canary \cite{canaryblog} \logmelspec is outperformed by \gammspec on all datasets. Interestingly, \logspec performs similar to \gammspec even though it has 5 times the dimensionality, which indicates that \gammspec retains most of the relevant information present in the raw spectrogram. The effectiveness of the gammatone filterbank is further evidenced by the fact that \gammfreqmask has much lower WER than \freqmask. Furthermore, while we expected \freqmask, \gammfreqmask and \diffgammspec to degrade WER because they discard some spectral information, we note that the degradation is minimal if any. In fact, on \librispeech test-clean both \diffgammspec and \gammfreqmask outperform \logmelspec under beam-search decoding. On \tedlium and \mlses, however, \diffgammspec has slightly higher WER than \logmelspec. As we shall see in the following sections, what \diffgammspec lacks in accuracy, it makes up for in robustness.

\subsection{Robustness to Adversarial Attacks}
\begin{table}
\small
    \begin{tabular}{llrr}
       Model & Feature  &  SNR & WER \\
       \midrule
       \multirow[c]{2}{*}{Branchformer} & DoGSpec & 13.60 & 11.03 \\
        & LogMelSpec & 25.10 & 5.15 \\
        \midrule
        \multirow[c]{9}{*}{Conformer} & DoGSpec & \textbf{8.60} & 0.00 \\
         & FreqMask & 10.80 & \textbf{19.12} \\
        & GammFreqMask & 10.10 & 7.35 \\
        & GammSpec & 15.20 & 13.24 \\
        & LogMelSpec & 21.30 & 6.62 \\
        & LogSpec & 15.00 & 3.68 \\
        & PNC & 13.40 & 8.09 \\
    \end{tabular}
    \label{tab:imp_adv}
    \caption{The WER between the prediction and target phrase for the various models and features and the SNR of the adversarially perturbed audio.}
\end{table}
\begin{figure}
    \centering
    \includegraphics[width=0.6\linewidth]{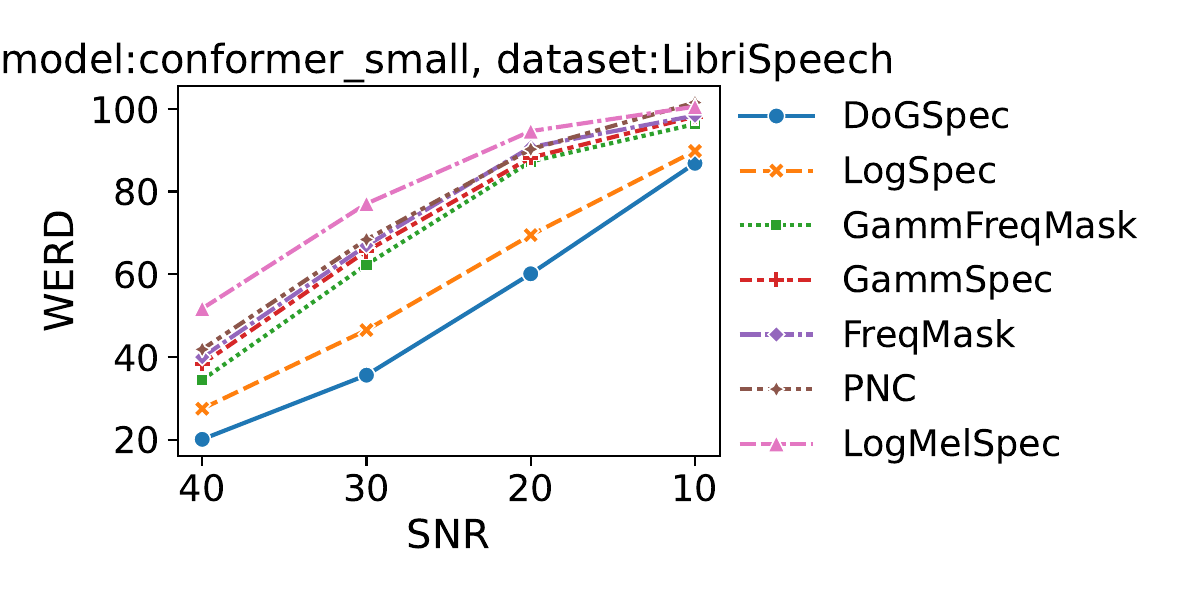}
    \includegraphics[width=\linewidth]{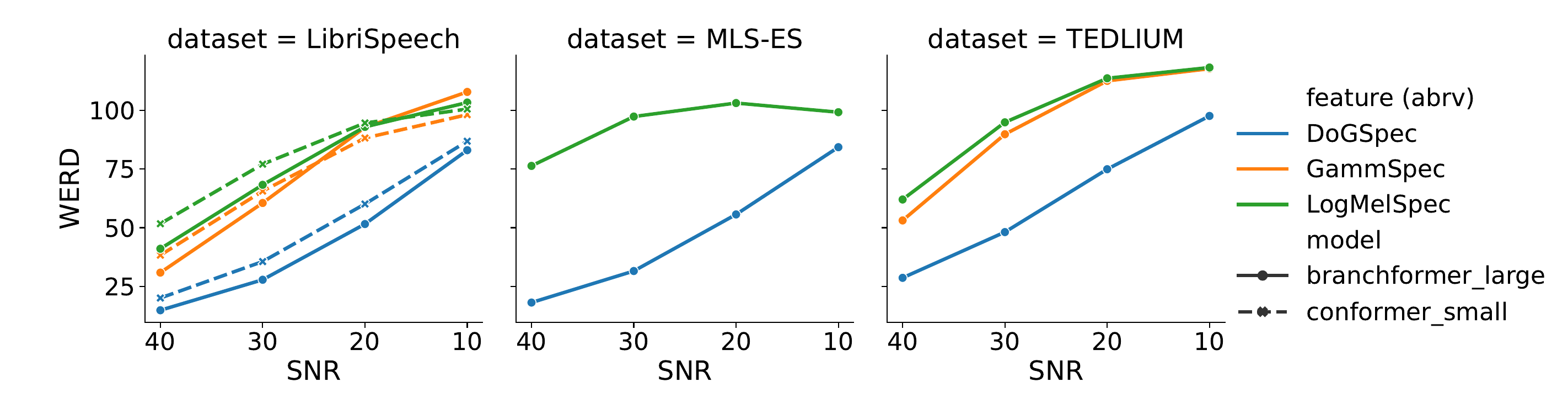}
    \caption{The WERD of models trained on various datasets and features against PGD attacks of increasing SNR bounds.}
    \label{fig:adv_werd}
\end{figure}
We consider two adversarial attacks in our evaluation: the untargetted PGD attack and targeted ``imperceptible'' attack. Figure \ref{fig:adv_werd} shows the WERD of the features and models against the PGD attack under different SNR bounds. We exclude \pncc and \mfcc from this analysis because they achieved very high WERs on clean data. We observe that \diffgammspec achieves the lowest WERD across all SNR bounds, followed by \logspec. Meanwhile, \logmelspec performs the worst. Interestingly, \freqmask, \gammfreqmask, \pnc and \gammspec perform similarly, and have much greater WERD than \diffgammspec and \logspec. Turning to Table \ref{tab:imp_adv}, we see that in terms of SNR \diffgammspec outperforms all models by forcing the attack to add much more noise to succeed. Under an SNR of 8.6 dB the ``imperceptible'' attack is going to be very perceptible and likely will significantly impact the intelligibility of speech, thus violating the basic requirements of an effective adversarial attack. Meanwhile, \logmelspec performs the worst a margin allowing the attack to achieve a low WER with 21.3 dB SNR, which may well be imperceptible, or barely perceptible. Interestingly, \freqmask has the highest WER without having the lowest WER, which may be because of gradient masking effects \cite{athalye2018obfuscated} due to discontinuities in the feature computation. We are surprised to observe that \pnc, despite being specifically designed to counteract noise, did not perform very well against adversarial attacks.

\subsection{Robustness to Non-Adversarial Noise}
\begin{table}[t!]
    \centering
    \small
    \begin{tabular}{p{1.1cm}llrr}
        Dataset & Model & Feature  & NWERD & WERD \\
        \midrule
        \multirow[c]{12}{*}{LibriSpeech} & \multirow[c]{3}{*}{Branchformer} & DoGSpec & 0.273 & 16.93 \\
        &  & GammSpec & \textbf{0.266} & 17.37 \\
        &  & LogMelSpec & \textbf{0.266} & \textbf{16.54} \\
\cline{2-5}        
& \multirow[c]{9}{*}{Conformer} & DoGSpec & 0.35 & 21.03 \\
        &  & FreqMask & 0.45 & 26.75 \\
        &  & GammFreqMask & 0.36 & 22.21 \\
        &  & GammSpec & \textbf{0.33} & \textbf{20.02} \\
        &  & LogMelSpec & 0.37 & 23.73 \\
        &  & LogSpec & 0.36 & 21.93 \\
        &  & MFCC & 0.58 & 33.84 \\
        &  & PNC & 0.38 & 22.84 \\
        &  & PNCC & 0.54 & 30.99 \\
        \hline
        \multirow[c]{3}{*}{TEDLIUM} & \multirow[c]{3}{*}{Branchformer} & DoGSpec & 0.53 & 30.72 \\
        &  & GammSpec & \textbf{0.48} & 29.42 \\
        &  & LogMelSpec & 0.49 & \textbf{28.89} \\
        \hline
        \multirow[c]{2}{*}{MLS-es} & \multirow[c]{2}{*}{Branchformer} & DoGSpec & 0.71 & 28.44 \\
        &  & LogMelSpec & \textbf{0.59} & \textbf{25.59} \\
    \end{tabular}
    \caption{WERD and NWERD of various models and features on the SRB benchmark.}
    \label{tab:srb_werd}
\end{table}
We evaluate the models on non-adversarial noisy speech recordings from the SRB benchmark and present aggregated values for NWERD and WERD in Table \ref{tab:srb_werd}. We see that for the conformer models trained on \librispeech and branchformer trained on \tedlium,\gammspec achieves the lowest NWERD, while the NWERD of \diffgammspec is slightly higher than \logmelspec. Interestingly, \pnc has one of the highest NWERD which indicates that the noise reduction mechanisms included in it do not generalize to diverse types of corruptions.
\section{Conclusion}
In this paper, we have revisited biologically plausible acoustic features for ASR. In this connection, we use several existing as well as our own novel biologically plausible acoustic features as inputs to modern transformer-based ASR systems and evaluate the robustness and accuracy of the resulting models. We find that using gammatone filterbank features improves both robustness and accuracy compared to the most popular mel filterbank features. Furthermore, the difference of gammatone spectrogram that we developed surpasses all other features in terms of adversarial robustness. Our work shows that the choice of feature has a significant influence on the accuracy and robustness of ASR models, and, thus, simply using the most popular feature of the times, such as mel filterbanks, may yield suboptimal outcomes. Moreover, we see that introducing biological mechanisms that are currently unrepresented in ASR DNNs can lead to improved robustness.

\bibliographystyle{unsrt}
\bibliography{refs}

\end{document}